\documentclass[pra,superscriptaddress,showpacs,nobalancelastpage]{revtex4}

\usepackage{graphicx}

\begin{document}

\title{ 
Macroscopic mechanical oscillators
at the quantum limit 
using optomechanical cooling}

\author{David Vitali}
\author{Stefano Mancini}
\affiliation{INFM, Dipartimento di Fisica,
Universit\`a di Camerino,
via Madonna delle Carceri, I-62032 Camerino, Italy}
\author{Luciano Ribichini} 
\affiliation{Albert Einstein Institut f\"ur Gravitationsphysik, Aussenstelle
Hannover, Callinstrasse 38, D-30167 Hannover, Germany}
\author{Paolo Tombesi}
\affiliation{INFM, Dipartimento di Fisica,
Universit\`a di Camerino,
via Madonna delle Carceri, I-62032 Camerino, Italy}

\begin{abstract}
We discuss how the optomechanical coupling provided by radiation 
pressure can be used to cool macroscopic collective degrees of 
freedom, as vibrational modes of movable mirrors. Cooling is achieved 
using a phase-sensitive feedback-loop which effectively overdamps the 
mirrors motion without increasing the thermal noise. 
Feedback results able to bring macroscopic objects 
down to the quantum limit. In particular, it is possible
to achieve squeezing and entanglement. 
\end{abstract}

\maketitle %% NULL FUNCTION WITH LATEX 2e

\section{Introduction}

Radiation pressure is a classical effect which is however at the basis
of laser cooling and of many techniques for the manipulation of 
quantum states of atoms. Its classical origin however 
naturally suggests to extend 
its application to a more macroscopic level. In fact, a number of 
recent papers \cite{MVTPRL,HEIPRL,PINARD,ATTO}
have shown how radiation pressure can be profitably used 
to cool macroscopic degrees of freedom as, for example, the 
vibrational modes of
a mirror of an optical cavity. The possibility to use optomechanical 
coupling to cool a cavity mirror in conjunction with a 
phase-sensitive feedback loop was first pointed out in 
Ref.~\cite{MVTPRL}. The optomechanical coupling allows one to
detect the mirror displacement through a homodyne measurement, 
and then the output photocurrent can be fed-back in such a way
to realize a real-time reduction of the mirror thermal fluctuations.
The scheme proposed in Ref.~\cite{MVTPRL} 
roughly amounts to a continuous version of the 
stochastic cooling technique used in accelerators \cite{ACC}. 
In fact, feedback continuously
``kicks'' the mirror in order to put it in its equilibrium position,
and for this reason the feedback scheme of Ref.~\cite{MVTPRL} 
has been called ``stochastic cooling feedback'' in \cite{LETTER,PRALONG}. 
However there are important differences with the scheme used in accelerators 
and its natural extension in the case of atomic clouds \cite{RAIZEN}.

A significant improvement has been then achieved
with the first experimental realization \cite{HEIPRL} 
of the feedback cooling scheme. The experiment however employed a 
slightly different feedback scheme, the so called
``cold damping'' technique which was already known in the study 
of classical electro-mechanical systems \cite{COLDD}. 
Cold damping amounts to applying a viscous feedback force 
to the oscillating mirror.
In the experiment of Ref.~\cite{HEIPRL}, 
the displacement of the mirror is measured with 
very high sensitivity, and the obtained information
is fed back to the 
mirror via the radiation pressure of another, intensity-modulated, laser
beam, incident on the back of the mirror.

The experiment of Ref.~\cite{HEIPRL} has been performed at room 
temperature, where the effects of quantum noise
are blurred by thermal noise and all the results can be well
explained in classical terms (see for example \cite{PINARD}). 
However, developing 
a fully quantum description of the system in the presence of feedback is
of fundamental importance, for two main reasons.
First of all it allows one to establish the conditions under which 
the effects of quantum noise in optomechanical systems become visible 
and experimentally detectable. For example, Ref.~\cite{LETTER}
has shown that there is an appreciable difference
between the classical and quantum description of feedback already 
at liquid He temperatures.
Moreover, a completely quantum treatment allows one to establish the ultimate
limits of the proposed feedback schemes, as for example, the 
possibility
to reach ground state cooling of a mechanical, macroscopic degree of 
freedom. Quantum limits of feedback cooling have
been already discussed in \cite{PRALONG,COURTY}, where the possibility
to reach ground state cooling has been shown for both schemes.
Here we shall review these results and make a detailed comparison
of the cooling capabilities of the two different schemes.  
Morever, the present paper will extend
the results of \cite{PRALONG} for what concerns the possibility
to achieve new nonclassical effects in the presence of feedback.
In particular we shall see that 
optomechanical cooling allows purely quantum effects, as
squeezing and entanglement, to come out 
also in macroscopic mechanical oscillators.
The experimental realization of these quantum 
limits in optomechanical systems is extremely difficult, but the 
feedback methods described in this paper may be useful also
for microelectromechanical systems, where the search for quantum 
effects in mechanical systems is also very active \cite{BLENC,ROUK,SCHWAB}.

The outline of the paper is as follows. In Sec. II 
we describe the model and derive the 
appropriate quantum Langevin equations. 
In Sec. III we describe both feedback schemes using the 
quantum Langevin theory developed in \cite{HOW,GTV}.
In Section IV we analyze the stationary state of the system, 
and we determine the cooling capabilities of both schemes.
Section V is devoted to the various nonclassical effects that can be
achieved with the feedback scheme of Ref.~\cite{MVTPRL},
while Section VI is for concluding remarks.

\section{The model}

\begin{figure}[ht]
\includegraphics[width=3.5in]{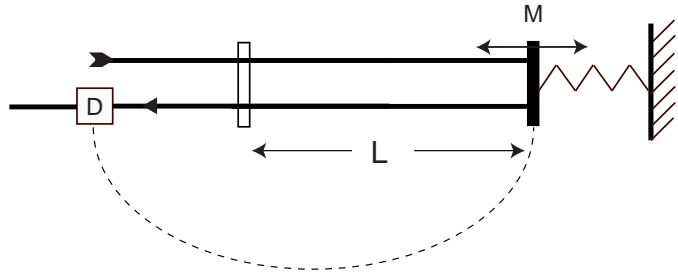}
\caption{Schematic description of a linear Fabry-Perot cavity
with the end oscillating mirror M.
The equilibrium cavity length is $L$.
A cavity mode is driven by an input laser beam.
The output field is subjected to homodyne detection (D).
The signal is then fed back to the mirror motion (dashed line). }
\label{fig1}
\end{figure}

\begin{figure}[ht]
\includegraphics[width=3.5in]{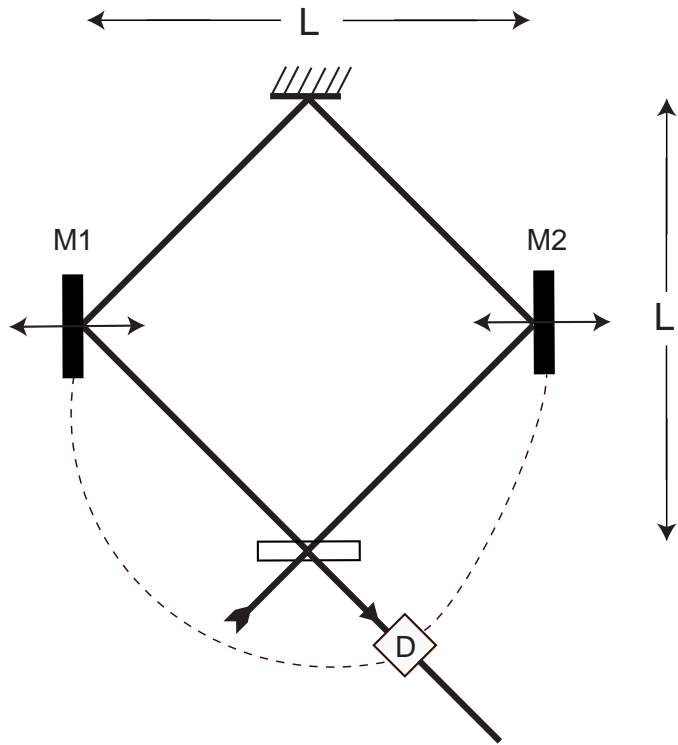}
\caption{Schematic description of a ring cavity
with two oscillating mirrors M1 and M2.
The equilibrium distance between them is $L$
(this is also the distance between the fixed mirrors).
A cavity mode is driven by an input laser beam.
The output field is subjected to homodyne detection (D).
The signal is then fed back to the mirrors motion (dashed lines).
}
\label{fig2}
\end{figure}

We shall consider two examples of optomechanical systems,
a Fabry-Perot cavity with an oscillating end mirror
(see Fig.~\ref{fig1} for a schematic description), and
a ring cavity with two oscillating 
mirrors (see Fig.~\ref{fig2}). 
These kind of optomechanical systems are used in high sensitivity
measurements, as the 
interferometric detection of gravitational waves \cite{GRAV}, and 
in atomic force microscopes \cite{AFM}. 
Their ultimate sensitivity is determined by the quantum fluctuations,
and therefore each movable 
mirror has to be described as a single {\em quantum} 
mechanical harmonic oscillator with mass $m$ and frequency 
$\omega_{m}$. The mirror motion is the result of the 
excitation of many vibrational modes, including internal acoustic modes.
The description of a movable mirror as a single oscillator is however a good
approximation when frequencies are limited to a bandwidth including 
a single mechanical resonance, by using for example a bandpass filter
in the detection loop \cite{hadjar2}.

The optomechanical coupling between the mirror and
the cavity field is realized by the radiation pressure. The 
electromagnetic field exerts a force on a movable mirror which is 
proportional to the intensity of the field, which, at the same time, 
is phase-shifted by an amount 
proportional to 
the mirror displacement from its equilibrium position.
In the adiabatic limit in which the mirror frequency is much smaller 
than the cavity free spectral range, 
one can focus on one cavity mode only, say of frequency
$\omega_{c}$, because
photon scattering into other modes can be neglected
\cite{LAW}. Moreover, in this adiabatic regime,
the generation of photons due to the Casimir effect,
and also retardation and Doppler effects are completely negligible
\cite{HAMI}.

Since we shall focus on the quantum and thermal noise of the system,
we shall neglect all the technical sources of noise, i.e., 
we shall assume that the driving laser is stabilized
in intensity and frequency. Including these 
supplementary noise sources is however quite straightforward and a 
detailed calculation of their effect is shown in Ref. \cite{KURT}. 
Moreover recent experiments have shown
that classical laser noise can be made negligible in the relevant 
frequency range \cite{HADJAR,TITTO}.

The dynamics of an optomechanical system is also influenced
by the dissipative interaction
with external degrees of freedom. The cavity mode is damped 
due to the
photon leakage through the mirrors which couple the cavity 
mode with the continuum of the outside electromagnetic modes. 
For simplicity 
we assume that the movable mirrors have perfect reflectivity and that
transmission takes place through a ``fixed'' mirror only.
Each mechanical oscillator, which may represent not only the
center-of-mass degree of freedom of the mirror, but also a torsional 
degree of freedom as in \cite{TITTO}, or an internal acoustic mode
as in \cite{HADJAR}, undergoes Brownian motion caused by the 
uncontrolled coupling with other internal and external modes at 
thermal equilibrium.

Let us first analyze the scheme of 
Fig.~\ref{fig1}, which can be described by the 
following Hamiltonian \cite{HAMI}
\begin{equation}
	H=\hbar \omega_{c}b^{\dagger}b + 
    \hbar\omega_m\left(P^{2}+
	Q^{2}\right) 
	-2\hbar G
	b^{\dagger}b Q + i\hbar E\left(b^{\dagger}e^{-i\omega_{0}t}-b
	e^{i\omega_{0}t}\right) \,, 
	\label{HINI}
\end{equation}
where $b$ is the cavity mode annihilation operator with optical
frequency $\omega_{c}$, and $E$ describes the coherent input field
with frequency $\omega_{0}\sim \omega_{c}$
driving the cavity.
The quantity $E$ is related to the input laser 
power $\wp$ by $E=\sqrt{\wp\gamma_{c}/\hbar \omega_{0}}$,
being $\gamma_c$ the photon decay rate.
Moreover, $Q$ and $P$ are the dimensionless position and momentum
operator of the movable mirror M. It is $\left[Q,P\right]=i/2$, 
and $G=(\omega_c/L)\sqrt{\hbar/2m\omega_m}$ represents the 
optomechanical coupling constant, with $L$
the equilibrium cavity length.

The dynamics of the system
can be described by the following set of coupled quantum 
Langevin equations (QLE)
(in the interaction picture with respect to 
$\hbar \omega_{0}b^{\dagger} b$) 
\begin{eqnarray}
\dot{Q}(t) &=& \omega_m P(t) \,, 
\label{QLENL1}\\
\dot{P}(t) &=& -\omega_{m} Q(t) + {\cal W}(t) -  
 {\gamma_m} P(t)   
 +G b^{\dagger}(t) b(t) \,,
\label{QLENL2}\\
\dot{b}(t) &=& - \left(i \omega_{c} - i \omega_{0} +
\frac{\gamma_{c}}{2}\right) b(t) + 2 i 
G Q(t) b(t) + E +
\sqrt{\gamma_{c}}b_{in}(t)\,, 
\label{QLENL3}
\end{eqnarray}
where $b_{in}(t)$ is the input noise operator \cite{GAR} 
associated with the vacuum 
fluctuations of the continuum of modes 
outside the cavity, having the 
following correlation functions 
\begin{eqnarray}
&& \langle b_{in}(t)b_{in}(t') \rangle = \langle 
b_{in}^{\dagger}(t)b_{in}(t') \rangle
= 0 \,,
\label{INCOR1}\\
&& \langle b_{in}(t)b_{in}^{\dagger}(t') \rangle = \delta(t-t') \,.
\label{INCOR2} 
\end{eqnarray}
Furthermore, ${\cal W}(t)$ is the quantum Langevin force acting on 
the mirror, with the following correlation function \cite{VICTOR},
\begin{equation}
\langle {\cal W}(t) {\cal W}(t^\prime) \rangle=
\frac{1}{2\pi}\frac{\gamma_m}{\omega_m} 
\Big\{ {\cal F}_{r}(t-t^\prime) + i 
{\cal F}_{i}(t-t^\prime) \Big\}\,,
\label{BROWCOR}
\end{equation}
where
\begin{eqnarray}
{\cal F}_{r}(t)&=&\int_{0}^{\varpi} d\omega \;
  \omega \cos(\omega t) \coth\left(\frac{\hbar\omega}{2 k_B T} 
  \right) \,,
  \label{FR} \\
{\cal F}_{i}(t)&=& - \int_{0}^{\varpi} d\omega \;
  \omega \sin(\omega t) \,,
  \label{FI}  
\end{eqnarray}
with $T$ the bath temperature, $\gamma_m$ the mechanical 
decay rate, $k_B$ the 
Boltzmann constant, and $\varpi$ the frequency cutoff of the reservoir
spectrum. Eqs.~(\ref{BROWCOR}), (\ref{FR}), and (\ref{FI})
show the non-Markovian nature of quantum Brownian motion, which 
becomes particularly evident in the low temperature limit 
\cite{GRAB,HAAKE}.
The symmetric correlation function 
becomes proportional to a Dirac delta function when the 
high temperature limit $ k_BT \gg \hbar \varpi $ first, 
and the infinite frequency 
cutoff limit $\varpi \rightarrow \infty$ later, are taken. 
It is only in this limit that the {\em exact} QLE
(\ref{QLENL1})-(\ref{QLENL3}) reduce to the standard ones
\cite{GAR}. The quantum Langevin description 
given by Eqs.~(\ref{QLENL1})-(\ref{QLENL3}) is more general than 
that associated with a master equation 
approach, which needs a high temperature assumption \cite{VICTOR}. 

In standard applications the driving field is
very intense so that the system is characterized by a
semiclassical steady state with 
the internal cavity mode in a coherent
state $|\beta\rangle $, and a new equilibrium position for the 
mirror, displaced by $G| \beta |^{2}/\omega_m$ 
with respect to that with no driving field.
The steady state amplitude $\beta $ is given by the solution
of the classical nonlinear equation 
$\beta=E/(\gamma_{c}/2+i \omega_{c} - i \omega_{0}
	+2 i G^2/\omega_{m}
        |\beta|^{2})$.
In this case, the dynamics is well described 
by linearizing the QLE 
(\ref{QLENL1})-(\ref{QLENL3}) around the steady state. Redefining
with $Q(t)$ and $b(t)$ the quantum 
fluctuations around the classical steady state, 
introducing the field phase 
$Y(t)=i\left(b^{\dagger}(t)-b(t)\right)/2$ and field amplitude 
$X(t)=\left(b(t)+b^{\dagger}(t)\right)/2$,  and choosing
the resulting cavity mode detuning 
$\Delta = \omega_{c} - \omega_{0} +
	2G^2/\omega_m \beta^{2}=0$
(by properly tuning the driving
field frequency $\omega_{0}$), the 
linearized QLEs can be rewritten as
\begin{eqnarray}
\dot{Q}(t) &=& \omega_m P(t) \,, 
\label{QLE2L1}\\
\dot{P}(t) &=& -\omega_{m} Q(t) -  
  {\gamma_m} P(t)   +2 G\beta X(t) + {\cal W}(t) \,,
\label{QLE2L2}\\
\dot{Y}(t) &=&  -\frac{\gamma_{c}}{2} Y(t)
+2 G \beta Q(t)+ \frac{\sqrt{\gamma_{c}}}{2} Y_{in}(t) \,,
\label{QLE2L3} \\
\dot{X}(t) &=&  -\frac{\gamma_{c}}{2} X(t)
+ \frac{\sqrt{\gamma_{c}}}{2} X_{in}(t) \,,
\label{QLE2L4}
\end{eqnarray}  
where we have introduced the phase input noise 
$Y_{in}(t)=i\left(b_{in}^{\dagger}(t)-b_{in}(t)\right)$ and the 
amplitude input noise $X_{in}(t)=b_{in}^{\dagger}(t)+b_{in}(t)$.

Similar arguments can be used to describe the dynamics of the
ring cavity scheme with two movable mirrors of Fig.~2. 
Assuming for simplicity that the two movable mirrors
are identical (i.e., equal mass, frequency and damping), and
since the two mirrors are perfectly reflecting, one can easily extend the QLEs
for the Fabry-Perot cavity (\ref{QLENL1})-(\ref{QLENL3}) to the
optomechanical system of Fig.~2, and get
(in the interaction picture with respect to 
$\hbar \omega_{0}b^{\dagger} b$) 
\begin{eqnarray}
    \dot{Q_{1}}(t) &=& \omega_m P_{1}(t) \,, 
    \label{QLENLQ1}\\
    \dot{P_{1}}(t) &=& -\omega_{m} Q_{1}(t) + {\cal W}_{1}(t) -  
    {\gamma_m} P_{1}(t)   
    +{\tilde G} b^{\dagger}(t) b(t) \,,
    \label{QLENLP1}\\
    \dot{Q_{2}}(t) &=& \omega_m P_{2}(t) \,, 
    \label{QLENLQ2}\\
    \dot{P_{2}}(t) &=& -\omega_{m} Q_{2}(t) + {\cal W}_{2}(t) -  
    {\gamma_m} P_{2}(t)   
    -{\tilde G} b^{\dagger}(t) b(t) \,,
    \label{QLENLP2}\\
    \dot{b}(t) &=& - \left(i \omega_{c} - i \omega_{0} +
    \frac{\gamma_{c}}{2}\right) b(t) + 2 i 
    {\tilde G} \left(Q_{1}(t) - Q_{2}(t)\right)b(t)+ E +
    \sqrt{\gamma_{c}}b_{in}(t)\,, 
    \label{QLENLb}
\end{eqnarray}
where $Q_{j}$ and $P_{j}$ $(j=1,\,2)$ 
are the dimensionless position and momentum
operators of the movable mirrors M1 and M2, with 
$\left[Q_{k},P_{j}\right]=(i/2)\delta_{kj}$, and 
${\tilde G}=(\omega_c/2\sqrt{2}L)\sqrt{\hbar/2m\omega_m}$ 
is the new coupling constant,
($L$ now represents the equilibrium distance between 
the movable mirrors, as well as the distance between the fixed 
mirrors, see Fig.~2).
The operators ${\cal W}_{1}$ and ${\cal W}_{2}$ are 
independent Brownian noise
operators acting on the mirrors M1 and M2 respectively, 
each of them with a correlation function specified by Eq.~(\ref{BROWCOR}).

It is convenient to
adopt the center of mass and relative distance coordinates and 
conjugate momenta,
$Q_{\pm}=(Q_{1}\pm Q_{2})/\sqrt{2}$, and 
$P_{\pm}=(P_{1}\pm P_{2})/\sqrt{2}$.
Then, Eqs.(\ref{QLENLQ1})-(\ref{QLENLb}) become
\begin{eqnarray}
    \dot{Q_{+}}(t) &=& \omega_m P_{+}(t) \,, 
    \label{QLENLQ+}\\
    \dot{P_{+}}(t) &=& -\omega_{m} Q_{+}(t) + {\cal W}_{+}(t)-  
    {\gamma_m} P_{+}(t)\,,
    \label{QLENLP+}\\
    \dot{Q_{-}}(t) &=& \omega_m P_{-}(t) \,, 
    \label{QLENLQ-}\\
    \dot{P_{-}}(t) &=& -\omega_{m} Q_{-}(t) + {\cal W}_{-}(t)-
    {\gamma_m} P_{-}(t)   
    +\sqrt{2}{\tilde G} b^{\dagger}(t) b(t) \,,
    \label{QLENLP-}\\
    \dot{b}(t) &=& - \left(i \omega_{c} - i \omega_{0} +
    \frac{\gamma_{c}}{2}\right) b(t) + 2 \sqrt{2} i 
    {\tilde G} Q_{-}(t) b(t) + E +
    \sqrt{\gamma_{c}}b_{in}(t)\,, 
    \label{QLENLbb}
\end{eqnarray}
where ${\cal W}_{\pm}=({\cal W}_{1}\pm{\cal W}_{2})/\sqrt{2}$
are again independent Brownian noise operators. 
It is evident from these equations
that only the relative motion of the two mrrors is coupled to the
radiation field, while the center of mass 
undergoes a simple Brownian motion.
Thus, neglecting the latter, the QLEs for the dynamics of the
relative motion and the cavity mode are identical to those of the
single mirror case, Eqs.~(\ref{QLENL1})-(\ref{QLENL3}), except for the
replacements $Q \to Q_{-}$, $P \to P_{-}$, and $G \to \sqrt{2}{\tilde G}$.
One can again consider an intensely driven cavity mode, and the fluctuations
around the semiclassical steady state, as discussed above
in the Fabry-Perot case. One has a consequently modified steady state
coherent amplitude $\tilde{\beta}$, and the linearized equations for
the relative motion and the cavity mode (assuming again
zero cavity mode detuning) become identical to those
of the Fabry-Perot case, Eqs.~(\ref{QLE2L1})-(\ref{QLE2L4}),
with the corresponding replacements
$Q \to Q_{-}$, $P \to P_{-}$, $\beta \to \tilde{\beta}$, 
and $G \to \sqrt{2}\tilde{G}$.

\section{Position measurement and feedback}

As it is shown by Eq.~(\ref{QLE2L3}),
when the driving and the cavity fields are resonant, the dynamics is 
simpler because only the phase quadrature $Y(t)$ is 
affected by the mirror position fluctuations $Q(t)$,
while the amplitude quadrature $X(t)$ is not.  
Therefore the mechanical motion
of the mirror can be detected by monitoring
the phase quadrature $Y(t)$. In the ring cavity case, the phase quadrature
$Y(t)$ instead reproduces the relative motion dynamics. 
However, to fix the ideas, 
we shall focus for simplicity onto the Fabry-Perot cavity case only.
The mirror position measurement is commonly performed in the 
large cavity bandwidth limit $\gamma_{c} \gg G \beta $, $\omega_m$,
when the 
cavity mode dynamics adiabatically follows that of the movable mirror
and it can be eliminated, that is, from Eq.~(\ref{QLE2L3}),
\begin{equation}
Y(t) \simeq \frac{4G \beta}{\gamma_{c}}Q(t) 
+\frac{Y_{in}(t)}{\sqrt{\gamma_{c}}},
\label{adiab}
\end{equation}
and $X(t) \simeq X_{in}(t)/\sqrt{\gamma_{c}}$ from Eq.~(\ref{QLE2L4}).
The experimentally detected quantity is the output homodyne photocurrent
\cite{HOW,GTV,homo}
\begin{equation}\label{BOUNDARY}
Y_{out}(t)=2\eta \sqrt{\gamma_{c}}Y(t)-\sqrt{\eta}Y_{in}^{\eta}(t)\,,
\end{equation}
where $\eta$ is the detection efficiency and $Y_{in}^{\eta}(t)$
is a generalized phase input noise, coinciding with the
input noise $Y_{in}(t)$ in the case of perfect detection $\eta 
=1$, and taking into account the additional noise due to the 
inefficient detection in the general case $\eta < 1$ \cite{GTV}.
This generalized phase input noise can be written in terms 
of a generalized input noise $b_{\eta}(t)$ as
$Y_{in}^{\eta}(t)=i\left[b_{\eta}^{\dagger}(t)-b_{\eta}(t)\right]$. 
The quantum noise $b_{\eta}(t)$ is correlated with the 
input noise $b_{in}(t)$ and it is characterized by the following 
correlation functions \cite{GTV}
\begin{eqnarray}
&& \langle b_{\eta}(t)b_{\eta}(t') \rangle = \langle 
b_{\eta}^{\dagger}(t)b_{\eta}(t') \rangle
= 0 \,,
\label{INCORETA1}\\
&& \langle b_{\eta}(t)b_{\eta}^{\dagger}(t') \rangle = \delta(t-t') 
\,,
\label{INCORETA2} \\
&& \langle b_{in}(t)b_{\eta}^{\dagger}(t')
\rangle = \langle b_{\eta}(t)b_{in}^{\dagger}(t')
\rangle = \sqrt{\eta}\delta(t-t').
\label{INCORETA3}  
\end{eqnarray}
The output of the homodyne measurement may be used to devise a
phase-sensitive feedback loop to control the dynamics of the mirror,
as it has been done in Ref.~\cite{MVTPRL}, 
and in Refs.~\cite{HEIPRL,PINARD} using cold damping.
Let us now see in detail how these two feedback schemes modify the 
quantum dynamics of the mirror.

In the scheme of Ref.~\cite{MVTPRL}, the feedback loop induces a 
continuous position shift controlled by the output homodyne 
photocurrent $Y_{out}(t)$.
This effect of feedback manifests itself in an additional term
in the QLE for a generic operator ${\cal O}(t)$ given by
\begin{equation}\label{DOSTRA}
\dot{{\cal O}}_{fb}(t)=
i\frac{\sqrt{\gamma_{c}}}{\eta}
\int_{0}^{t} dt'G_{mf}(t')Y_{out}(t-t')
\left[g_{mf}P(t),{\cal O}(t)\right]\,,
\end{equation}
where $G_{mf}(t)$ is the feedback transfer function, and $g_{mf}$ is a 
feedback gain factor.  The implementation of this scheme is nontrivial
because it is equivalent to add a feedback interaction linear in the mirror 
momentum, as it could be obtained with a charged mirror
in a homogeneous magnetic field. For this reason here we shall refer to it
as ``momentum feedback'' (see, however, the recent parametric 
cooling scheme demonstrated in Ref.~\cite{ATTO}, showing some 
similarity with the feedback scheme of Ref.~\cite{MVTPRL}).

Feedback is characterized by a delay time which is essentially determined 
by the electronics and is always much smaller than the typical 
timescale of the mirror dynamics. It is therefore common to consider
the zero delay-time limit $G_{mf}(t) \sim \delta(t)$, which is quite delicate
in general \cite{HOW,GTV}. However, for linearized systems, the limit
can be taken directly in Eq.~(\ref{DOSTRA}) \cite{PRALONG}, 
so to get the following QLE in the presence of feedback 
\begin{eqnarray}
\dot{Q}(t)&=&\omega_{m}P(t)
+g_{mf}\gamma_{c}Y(t)-\frac{g_{mf}}{2}\sqrt{
\frac{\gamma_c}{\eta}}Y_{in}^{\eta}(t)\,,
\label{QFBEQ1}\\
\dot{P}(t) &=& -\omega_{m} Q(t) -  
  {\gamma_m} P(t)   +2 G\beta X(t) + {\cal W}(t)\,,
\label{QFBEQ2}\\
\dot{Y}(t) &=&  -\frac{\gamma_{c}}{2} Y(t)
+2 G \beta Q(t)+ \frac{\sqrt{\gamma_{c}}}{2} Y_{in}(t) \,,
\label{QFBEQ3} \\
\dot{X}(t) &=&  -\frac{\gamma_{c}}{2} X(t)
+ \frac{\sqrt{\gamma_{c}}}{2} X_{in}(t) \,,
\label{QFBEQ4}
\end{eqnarray} 
where we have used Eq.~(\ref{BOUNDARY}).
After the adiabatic elimination of the radiation mode (see 
Eq.~(\ref{adiab})), and introducing the rescaled, dimensionless, input 
power of the driving laser 
\begin{equation} 
\zeta =  \frac{16 G^{2}\beta ^{2}}{\gamma_m\gamma_{c}}=
\frac{64G^{2}}{\hbar \omega_{0}\gamma_{m}\gamma_{c}^{2}}\wp ,
\label{zeta}
\end{equation}  
and the rescaled feedback gain
$g_1 = -4G\beta g_{mf}/\gamma_{m}$, 
the above equations reduce to
\begin{eqnarray}
\dot{Q}(t)&=&\omega_{m}P(t)
-\gamma_m g_{1} Q(t)-\sqrt{\frac{\gamma_{m}}{\zeta}}g_{1}
Y_{in}(t)+\sqrt{\frac{\gamma_{m}}{\eta \zeta}}\frac{g_{1}}{2}
Y_{in}^{\eta}(t)\,,
\label{QEQ1}\\
\dot{P}(t)&=&-\omega_{m}Q(t)
-\gamma_{m}P(t)+\frac{1}{2}\sqrt{\gamma_{m}\zeta}
X_{in}(t)+{\cal W}(t)\,.
\label{QEQ2}
\end{eqnarray}
This treatment explicitly includes the limitations due to the 
quantum efficiency of the detection, but neglects other
possible technical imperfections of the feedback loop, as for
example the electronic noise of the feedback loop, whose effects have 
been discussed in \cite{PINARD}.

Cold damping techniques
have been applied in classical electromechanical systems
for many years \cite{COLDD}, 
and only recently they have been proposed to improve
cooling and sensitivity at the quantum level \cite{GRAS}.
This technique is based on the application of a negative derivative
feedback, which increases the damping of the system without 
correspondingly increasing the thermal noise \cite{COLDD,GRAS}.
This technique has been succesfully applied for the first
time to an optomechanical 
system composed of a high-finesse cavity with a movable mirror in 
the experiments of Refs.~\cite{HEIPRL,PINARD,ATTO}. 
In these experiments, the displacement of the mirror is measured with 
very high sensitivity \cite{ATTO,HADJAR}, and the obtained information
is fed back to the 
mirror via the radiation pressure of another, intensity-modulated, laser
beam, incident on the back of the mirror.
Cold damping is obtained by modulating with the {\em time derivative}
of the homodyne signal, in such a way that
the radiation 
pressure force is proportional to the mirror velocity.
The results of Refs.~\cite{HEIPRL,PINARD,ATTO} referred to a room temperature
experiment, and have been explained using a classical description.
The quantum description of cold damping
has been instead presented in \cite{GRAS} using quantum network theory, and
in \cite{PRALONG,LETTER} using a quantum Langevin description.
In this latter treatment, cold damping 
implies the following additional term
in the QLE for a generic operator ${\cal O}(t)$,
\begin{equation}\label{DLORO}
\dot{{\cal O}}_{fb}(t)=
\frac{i}{\eta \sqrt{\gamma_{c}}}
\int_{0}^{t} dt'G_{cd}(t')Y_{out}(t-t')
\left[g_{cd}Q(t),{\cal O}(t)\right]\,.
\end{equation}
As in the previous case, one usually assume a Markovian feedback loop 
with negligible delay. Since one needs a derivative feedback, this 
would ideally imply $G_{cd}(t)= -\delta'(t)$, i.e., 
$\tilde{G}_{cd}(\omega) = i\omega $, $\forall \omega $, even though, 
in practice, it is sufficient to satisfy this 
condition within the detection bandwidth $\Delta \omega $.
In this case, the QLEs for the cold damping 
feedback scheme become
\begin{eqnarray}
\dot{Q}(t)&=&\omega_{m}P(t) \,,
\label{CFBEQ1}\\
\dot{P}(t) &=& -\omega_{m} Q(t) -  
  {\gamma_m} P(t)   +2 G\beta X(t)  
  -g_{cd}\dot{Y}(t)+\frac{g_{cd}}{2\sqrt{\gamma_c \eta}}
  \dot{Y}_{in}^{\eta}(t)
+{\cal W}(t)\,,
\label{CFBEQ2}\\
\dot{Y}(t) &=&  -\frac{\gamma_{c}}{2} Y(t)
+2 G \beta Q(t)+ \frac{\sqrt{\gamma_{c}}}{2} Y_{in}(t) \,,
\label{CFBEQ3} \\
\dot{X}(t) &=&  -\frac{\gamma_{c}}{2} X(t)
+ \frac{\sqrt{\gamma_{c}}}{2} X_{in}(t) \,.
\label{CFBEQ4}
\end{eqnarray} 
Adiabatically 
eliminating the cavity mode, and introducing the 
rescaled, dimensionless feedback gain 
$g_{2}=4 G \beta \omega_{m}g_{cd}/\gamma_{m}\gamma_{c}$,
one has
\begin{eqnarray}
&& \dot{Q}(t) = \omega_m P(t), 
 \label{QLEFCDAD1}\\
&&\dot{P}(t) = -\omega_{m} Q(t) -  
  {\gamma_m} P(t)   +\frac{1}{2}\sqrt{\gamma_{m}\zeta}
  X_{in}(t) + {\cal W}(t)  \nonumber \\
&&  - \frac{\gamma_m g_{2}}
 {\omega_m}\dot{Q}(t)-\frac{g_{2}\sqrt{\gamma_{m}}}{\omega_m \sqrt{\zeta}}
 \dot{Y}_{in}(t)
 + \frac{g_{2}\sqrt{\gamma_{m}}}{2\omega_m \sqrt{\eta \zeta}}
 \dot{Y}_{in}^{\eta}(t).
\label{QLEFCDAD2}
\end{eqnarray}
The presence of an ideal derivative feedback
implies the introduction of two new quantum input noises,
$\dot{Y}_{in}(t)$ and $\dot{Y}_{in}^{\eta}(t)$, whose correlation functions
can be simply obtained by differentiating the corresponding correlation
functions of $Y_{in}(t)$ and $Y_{in}^{\eta}(t)$. We have therefore
\begin{eqnarray}
\langle \dot{Y}_{in}(t) \dot{Y}_{in}(t')\rangle &=&
\langle \dot{Y}_{in}(t') \dot{Y}_{in}(t)\rangle = 
\langle \dot{Y}_{in}^{\eta}(t) \dot{Y}_{in}^{\eta}(t')\rangle =
\langle \dot{Y}_{in}^{\eta}(t') \dot{Y}_{in}^{\eta}(t)\rangle
=-\ddot{\delta}(t-t'), \label{dotcorre1}\\
 \langle \dot{Y}_{in}^{\eta}(t) \dot{Y}_{in}(t')\rangle &=&
\langle \dot{Y}_{in}(t') \dot{Y}_{in}^{\eta}(t)\rangle
=-\sqrt{\eta}\ddot{\delta}(t-t'), \label{dotcorre2}\\
\langle X_{in}(t) \dot{Y}_{in}^{\eta}(t')\rangle &=&
-\langle \dot{Y}_{in}^{\eta}(t') X_{in}(t)\rangle
=-i\sqrt{\eta}\dot{\delta}(t-t'), \label{dotcorre3}
\end{eqnarray}
which however, as discussed above, have to be considered as 
approximate expressions valid within the detection bandwidth only.

The two sets of QLE for the mirror Heisenberg operators,
Eqs.~(\ref{QEQ1})-(\ref{QEQ2}) and (\ref{QLEFCDAD1})-(\ref{QLEFCDAD2}), 
show that the two feedback 
schemes are not exactly equivalent. They are however physically 
analogous, as it can be seen, for example, by looking at the 
differential equation for the displacement operator $Q(t)$. 
In fact, from Eqs.~(\ref{QEQ1}) and (\ref{QEQ2}) one gets 
\begin{eqnarray}
&&\ddot{Q}(t)+\left(1+g_{1}\right)\gamma_{m}\dot{Q}(t) 
+\left(\omega_{m}^{2}+\gamma_{m}^{2}g_{1}\right)Q(t) \nonumber \\
&&=\omega_{m}\left[
\frac{1}{2}\sqrt{\gamma_{m}\zeta}X_{in}(t)+{\cal W}(t)\right]
-\sqrt{\frac{\gamma_{m}}{\zeta}}g_{1}
\left(\dot{Y}_{in}(t)+\gamma_m Y_{in}(t)\right) \\
&&+\sqrt{\frac{\gamma_{m}}{\eta \zeta}}\frac{g_{1}}{2}
\left(\dot{Y}_{in}^{\eta}(t)+\gamma_m Y_{in}^{\eta}(t)
\right) ,\nonumber 
\end{eqnarray}
for the momentum feedback scheme,
while from Eqs.~(\ref{QLEFCDAD1}) and (\ref{QLEFCDAD2}) one gets
\begin{eqnarray}
&&\ddot{Q}(t)+\left(1+g_{2}\right)\gamma_{m}\dot{Q}(t) +
\omega_{m}^{2}Q(t) \\
&& =\omega_{m}\left[\frac{1}{2}\sqrt{\gamma_{m}\zeta}
  X_{in}(t) + {\cal W}(t) -\frac{g_{2}\sqrt{\gamma_{m}}}{\omega_m 
  \sqrt{\zeta}}\dot{Y}_{in}(t)
 + \frac{g_{2}\sqrt{\gamma_{m}}}{2\omega_m 
  \sqrt{\eta \zeta}}\dot{Y}_{in}^{\eta}(t)
\right], \nonumber
\end{eqnarray}
for the cold damping scheme. 
The comparison shows that in both schemes the main effect of 
feedback is the modification of mechanical damping $\gamma_{m} 
\rightarrow \gamma_{m}(1+g_{i})$ ($i=1,2$). In the momentum feedback
scheme one 
has also a frequency renormalization $\omega_{m}^{2} \rightarrow  
\omega_{m}^{2}+\gamma_{m}^{2}g_{1}$, which is however negligible 
when the mechanical quality factor ${\cal Q}=\omega_{m}/\gamma_{m}$
is large. Moreover, the noise terms are similar, but not identical.
In particular, the feedback scheme of Ref.~\cite{MVTPRL}
is also subject to the phase noises $Y_{in}(t)$ and  
$Y_{in}^{\eta}(t)$, while cold damping is not. However, 
the comparison shows that also momentum 
feedback provides a cold 
damping effect of increased damping without an increased temperature.

\section{Cooling and stationary state}

We now study the stationary state of the movable mirror in the presence 
of both feedback schemes, which is obtained by considering the 
dynamics in the asymptotic limit $t \to \infty$. 
We shall see 
that, in both cases, ground state cooling can be achieved.

\subsection{Cold damping}

Now we characterize the stationary state of the mirror in the 
presence of cold damping. This stationary state has been already studied
using classical arguments in \cite{HEIPRL,PINARD}, while the discussion of
the cooling limits of cold damping in the quantum case has been 
recently presented in \cite{COURTY}. The results
of \cite{COURTY} have been generalized to the case of 
nonideal quantum efficiency $\eta < 1$ in \cite{PRALONG}, and here we 
shall review these results and compare 
in detail the cooling capabilities of the two feedback schemes.

The stationary variances can be expressed in terms of the Fourier
transforms of the noise correlation functions, and using 
the correlation functions 
(\ref{INCOR1}), (\ref{INCOR2}), (\ref{BROWCOR}), 
(\ref{INCORETA1})-(\ref{INCORETA3}), and 
(\ref{dotcorre1})-(\ref{dotcorre3}), one can write
\begin{eqnarray}
\langle Q^{2}\rangle _{st} &=& \gamma_{m}
\int_{-\infty}^{\infty}\frac{d \omega}{2 
\pi}\left|{\tilde \chi}_{cd}(\omega)\right|^{2}\left[\frac{\zeta}{4}
+\frac{g_2^2}{4 \eta \zeta }
\frac{|\tilde{G}_{cd}(\omega)|^{2}}{\omega_m^2} \right. \nonumber \\
&& \left.+\frac{\omega}{2 \omega_{m}} \coth
\left(\frac{\hbar \omega}{2 k_{B} T}\right)\Theta_{[-\varpi,\varpi]}(\omega) 
\right],
\label{q2fou} \\
\langle P^{2}\rangle _{st} &=& \gamma_{m}
\int_{-\infty}^{\infty}\frac{d \omega}{2 
\pi} \frac{\omega^2}{\omega_m^2}\left 
|{\tilde \chi}_{cd}(\omega)\right|^{2}\left[\frac{\zeta}{4}
+\frac{g_2^2}{4 \eta \zeta}
\frac{|\tilde{G}_{cd}(\omega)|^{2}}{\omega_m^2} \right. \nonumber \\
&& \left. +\frac{\omega}{2 \omega_{m}} \coth
\left(\frac{\hbar \omega}{2 k_{B} T}\right)\Theta_{[-\varpi,\varpi]}(\omega)
\right] ,
\label{p2fou} 
\end{eqnarray} 
where 
\begin{equation}
{\tilde \chi}_{cd}(\omega)=\frac{\omega_{m}}{\omega_{m}^{2}
	-\omega^{2}+i\omega \gamma_{m}\left(1+g_{2}\right)}
	\label{susccd}
\end{equation}
is the frequency-dependent susceptibility of the mirror in the 
cold damping feedback scheme, and 
$\Theta_{[-\varpi,\varpi]}(\omega)$ is a gate function equal to 1 
for $|\omega | < \varpi $ and zero otherwise. 

In general, each steady state variance has three contributions: 
i) the back action of the radiation pressure, 
proportional to the input power $\zeta$; ii) the feedback-induced noise term
proportional to the feedback gain squared,
and inversely proportional to the input 
power; iii)  the thermal noise term due to the 
mirror Brownian motion.
The back-action contribution can be evaluated in a straightforward way 
for both variances, and it is given by
\begin{equation}
\langle Q^{2}\rangle _{st}^{ba}=\langle P^{2}\rangle _{st}^{ba}=
\frac{\zeta}{8(1+g_{2})}. \label{pba}
\end{equation}
The feedback-induced contribution generally depends upon the form of
$|\tilde{G}_{cd}(\omega)|^{2}$, where $\tilde{G}_{cd}(\omega)$
is a causal function (i.e., it is analytic for ${\rm Im}\omega < 0$) and 
approximatively equal to $i\omega $ within the detection bandwidth 
$\Delta \omega $. The factor $|{\tilde \chi}_{cd}(\omega)|^2$ 
is highly peaked around the mechanical resonance $\omega =\omega_m$, 
with width $\gamma_m(1+g_2)$. Since it is usually
$\gamma_m(1+g_2) < \Delta \omega$,  one can safely approximate 
$|\tilde{G}_{cd}(\omega)|^{2}$ with its value at the 
resonance peak, $|\tilde{G}_{cd}(\omega)|^{2}\simeq \omega_{m}^{2}$, and obtain
\begin{equation}
\langle Q^{2}\rangle _{st}^{fb}=\langle P^{2}\rangle _{st}^{fb}=
\frac{g_{2}^{2}}{8\eta \zeta (1+g_{2})}. \label{pfb}
\end{equation}
The Brownian motion contribution is generally cumbersome
and it has been already exactly evaluated for both variances
in \cite{GRAB}. However, in optomechanical systems, 
the condition $\hbar \omega_m \ll k_B T$ is commonly met, and the classical
approximation $\coth(\hbar \omega /2k_B T) \simeq 2k_B T/\hbar \omega$
can be made in the expression for $\langle Q^{2}\rangle _{st}$,
obtaining 
\begin{equation}
\langle Q^{2}\rangle _{st}^{BM}=
\frac{k_B T}{2\hbar \omega_{m}(1+g_{2})}. \label{qbm}
\end{equation}
In the evaluation of $\langle P^{2}\rangle _{st}^{BM}$ instead, 
the classical approximation 
has to be made with 
care, because, due to the presence of the $\omega^{2}$ term, 
the integral (\ref{p2fou}) has an ultraviolet divergence in the usually 
considered $\varpi \to \infty $ limit (see also Eq.~(\ref{susccd})).
This means that, differently from $\langle Q^{2}\rangle_{st}^{BM}$,
the classical approximation for $\langle P^{2}\rangle_{st}^{BM}$ 
is valid only under the {\em stronger} condition $\hbar 
\varpi \ll k_B T $ \cite{GRAB}, 
and that in the intermediate temperature range
$ \hbar \varpi \gg k_B T \gg \hbar \omega_{m}$ (which may be of interest
for optomechanical systems), one has an additional
logarithmic correction, so to get
\begin{equation}
\langle P^{2}\rangle _{st}^{BM}=
\frac{k_B T}{2\hbar \omega_{m}}
\frac{1}{1+g_{2}}+
\frac{\gamma_{m}}{\pi \omega_{m}}\ln\left(\frac{\hbar \varpi}{2 \pi k 
T}\right). \label{pbm3}
\end{equation}
However, in the common situation of a high ${\cal Q}$ mechanical mode, 
this logarithmic correction can be neglected, the
dependence on the frequency cutoff $\varpi$ vanishes, and one finally gets
\begin{equation}
\langle Q^{2}\rangle _{st} = \langle P^{2}\rangle _{st} =
\left[\frac{g_{2}^{2}}{8 \eta \zeta } +
\frac{\zeta }{8}+
\frac{k_B T}{2\hbar \omega_{m}}\right]
\frac{1}{1+g_{2}}.
\label{q2cd}
\end{equation}
This fact, together with the fact that
\begin{equation}
\langle PQ+QP\rangle _{st} = \frac{1}{\omega_m}\lim_{t \to \infty}
\frac{d}{dt}\langle Q(t)^{2}\rangle =0, 
\end{equation}
implies that the stationary state in the presence
of cold damping is an effective thermal state
with a mean excitation number $\langle n \rangle = 2 
\langle Q^{2}\rangle _{st} -1/2$, where $\langle Q^{2}\rangle _{st}$ is given
by Eq.~(\ref{q2cd}). This effective thermal equilibrium state
in the presence of cold damping has been already pointed out in 
\cite{HEIPRL,PINARD}, within a classical treatment neglecting both the 
back-action and the feedback-induced terms. The present fully quantum analysis
shows that cold damping has two opposite effects on the effective equilibrium
temperature of the mechanical mode: on one hand $T$ is reduced by the factor
$(1 +g_2)^{-1}$, but, on the other hand, the effective temperature
is increased by the additional noise terms.

Let us now consider the optimal conditions for 
cooling, and the cooling limits of the cold damping feedback scheme.
Neglecting
the logarithmic correction to $\langle P^{2}\rangle _{st}^{BM}$,
the stationary oscillator energy is given by
\begin{equation}
U_{st} = 2\hbar \omega_m \langle Q^{2}\rangle _{st} = 
\frac{\hbar \omega_m}{4\left(1
+g_{2}\right)}\left[
\frac{g_{2}^{2}}{\eta \zeta }+
\zeta+ \frac{4k_B T}{\hbar \omega_{m}}\right].
\label{ener2}
\end{equation} 
This expression coincides with that derived and discussed
in \cite{COURTY}, except for the
presence of the homodyne detection efficiency $\eta$, which was
ideally assumed equal to one in \cite{COURTY}. The optimal conditions
for cooling can be derived in the same way as it has been done in
\cite{COURTY}. The energy $U_{st}$ is minimized with
respect to $\zeta$ keeping $g_2$ fixed, 
thereby getting $\zeta_{opt} = g_2/\sqrt{\eta}$. Under these 
conditions, the stationary oscillator energy becomes
\begin{equation}
U_{st} = \frac{\hbar \omega_m}{2}\frac{g_2}{1+g_2}
\left[\frac{1}{\sqrt{\eta}}+
\frac{2k_B T}{\hbar \omega_m}\frac{1}{g_2}\right]\;,
\label{enersimpl2}
\end{equation}
showing that, in the ideal limit 
$\eta =1$, $g_2 \to \infty$ (and therefore $\zeta \sim g_2 \to \infty$), 
cold damping is able to reach the
quantum limit $U_{st} =\hbar \omega_m /2$, i.e., it is able to cool
the mirror to its quantum ground state, as first pointed out
in \cite{COURTY}. 

\begin{figure}[ht]
\includegraphics[width=3.5in]{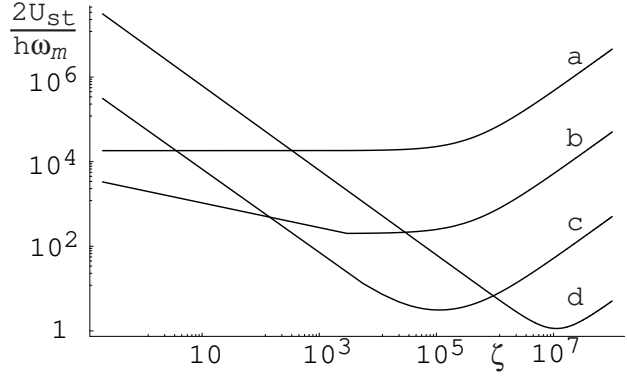}
\caption{Rescaled steady-state energy $2U_{st}/\hbar \omega_{m}$ 
versus the rescaled input power $\zeta$, plotted for 
different values of $g_{2}$ (a: $g_{2}=10$, b: $g_{2}=10^{3}$,
c: $g_{2}=10^{5}$, d: $g_{2}=10^{7}$),
with $k_{B}T/\hbar \omega_{m}=10^{5}$ and $\eta =0.8$.
The optimal input power correspondingly increases, and
for high gain values, ground state cooling can be 
obtained.}
\label{enecd-g}
\end{figure}

Fig.~\ref{enecd-g} shows the rescaled steady-state energy
$2U_{st}/\hbar \omega_{m}$ versus $\zeta$ plotted for 
increasing values of $g_{2}$ (a: $g_{2}=10$, b: $g_{2}=10^{3}$,
c: $g_{2}=10^{5}$, d: $g_{2}=10^{7}$),
with $k_{B}T/\hbar \omega_{m}= 10^{5}$ and $\eta =0.8$.
For high gain values, ground state cooling can be 
achieved, even with nonunit homodyne detection 
efficiency. 

\subsection{Momentum feedback}

One can proceed in a similar manner for the evaluation of the 
steady-state variances in the case of the feedback of 
Ref.~\cite{MVTPRL}. One has again the back-action, 
the feedback-induced, and the 
Brownian motion contributions. Differently from cold damping 
however, the variances now strongly depend on the mechanical quality 
factor ${\cal Q}$ and each noise contributes differently to 
$\langle Q^{2}\rangle _{st}$ and $\langle P^{2}\rangle _{st}$.
In the Markovian, zero-delay limit it is $\tilde{G}_{mf}(\omega)=1$, 
and the feedback-induced noise is delta-correlated as the 
back-action noise. Using the correlation functions 
(\ref{INCOR1}), (\ref{INCOR2}), (\ref{BROWCOR}), and 
(\ref{INCORETA1})-(\ref{INCORETA3}), one arrives at \cite{PRALONG}
\begin{eqnarray}
\langle Q^{2}\rangle _{st}^{ba} &=& 
\frac{\zeta {\cal Q}^{2}}{8\left(1+g_{1}\right)
\left({\cal Q}^{2}+g_{1}\right)},
\label{q2sc2ba} \\
\langle P^{2}\rangle _{st}^{ba} &=& 
\frac{\zeta \left({\cal Q}^{2}+g_{1}^{2}+g_{1}\right)}{8\left(1+g_{1}\right)
\left({\cal Q}^{2}+g_{1}\right)},
\label{p2sc2ba}
\end{eqnarray}
for the back-action contribution, and
\begin{eqnarray}
\langle Q^{2}\rangle _{st}^{fb} &=& 
\frac{g_{1}^{2}}{8 \eta \zeta} 
\frac{1+{\cal Q}^{2}+g_{1}}{\left(1
+g_{1}\right)\left({\cal Q}^{2}+g_{1}\right)},
\label{q2sc2fb} \\
\langle P^{2}\rangle _{st}^{fb} &=& 
\frac{g_{1}^{2}}{8 \eta \zeta} 
\frac{{\cal Q}^{2}}{\left(1
+g_{1}\right)\left({\cal Q}^{2}+g_{1}\right)},
\label{p2sc2fb}
\end{eqnarray}
for the feedback-induced contribution.

Finally, for the Brownian motion contribution, one has a situation analogous
to that discussed for the cold damping scheme.
The generally valid expressions for the variances
have been already exactly evaluated in
\cite{GRAB}. However, in the limit $\hbar \omega_m \ll k_B T$,
we can make the classical
approximation $\coth(\hbar \omega /2k_B T) \simeq 2k_B T/\hbar \omega$
in the expression for $\langle Q^{2}\rangle _{st}$,
obtaining 
\begin{equation}
\langle Q^{2}\rangle _{st}^{BM}=
\frac{k_B T}{2\hbar \omega_{m}}\frac{{\cal Q}^2}{(1+g_{2})({\cal Q}^2+g_1)}. 
\label{qbmsc}
\end{equation}
Also in this feedback scheme, the Brownian motion contribution
to $\langle P^{2}\rangle _{st}$ has an ultraviolet divergence in the usually 
considered $\varpi \to \infty $ limit \cite{PRALONG}.
Again, the classical approximation for $\langle P^{2}\rangle_{st}^{BM}$ 
is valid only under the {\em stronger} condition $\hbar 
\varpi \ll k_B T $ \cite{GRAB}, 
and, in the intermediate temperature range
$ \hbar \varpi \gg k_B T \gg \hbar \omega_{m}$, one has an additional
logarithmic correction, i.e.,
\begin{equation}
\langle P^{2}\rangle _{BM} = \frac{k_B T}{2\hbar \omega_{m}}
\frac{g_{1}^{2}+{\cal Q}^{2}+g_{1}}{\left(1
+g_{1}\right)\left({\cal Q}^{2}+g_{1}\right)}+
\frac{\gamma_{m}}{\pi \omega_{m}}\ln\left(\frac{\hbar \varpi}{2 \pi k 
T}\right). \label{pbm2}
\end{equation}
Summing up the three contributions, we arrive at \cite{PRALONG}
\begin{eqnarray}
\langle Q^{2}\rangle _{st} &=& 
\frac{g_{1}^{2}}{8 \eta \zeta} 
\frac{1+{\cal Q}^{2}+g_{1}}{\left(1
+g_{1}\right)\left({\cal Q}^{2}+g_{1}\right)}+
\left[\frac{\zeta}{8}+
\frac{k_B T}{2\hbar \omega_{m}}\right]
\frac{{\cal Q}^{2}}{\left(1+g_{1}\right)
\left({\cal Q}^{2}+g_{1}\right)},
\label{q2sc2} \\
\langle P^{2}\rangle _{st} &=& 
\frac{g_{1}^{2}}{8 \eta \zeta}
\frac{{\cal Q}^{2}}{\left(1+g_{1}\right)
\left({\cal Q}^{2}+g_{1}\right)}+
\left[\frac{\zeta}{8}+
\frac{k_B T}{2\hbar \omega_{m}}\right]
\frac{g_{1}^{2}+{\cal Q}^{2}+g_{1}}{\left(1
+g_{1}\right)\left({\cal Q}^{2}+g_{1}\right)} \nonumber \\
&&+ \frac{\gamma_{m}}{\pi \omega_{m}}\ln\left(\frac{\hbar \varpi}{2 \pi k 
T}\right).
\label{p2sc2}
\end{eqnarray} 
These expressions coincide with the corresponding ones obtained in 
\cite{MVTPRL} using a Master equation description, except for the 
logarithmic correction for $\langle P^{2}\rangle _{st}$, 
which however, in the case of mirror with 
a good quality factor ${\cal Q}$, is quite 
small, even in the intermediate
temperature range $ \hbar \varpi \gg k_B T \gg \hbar \omega_{m}$.

\begin{figure}[ht]
\includegraphics[width=3.5in]{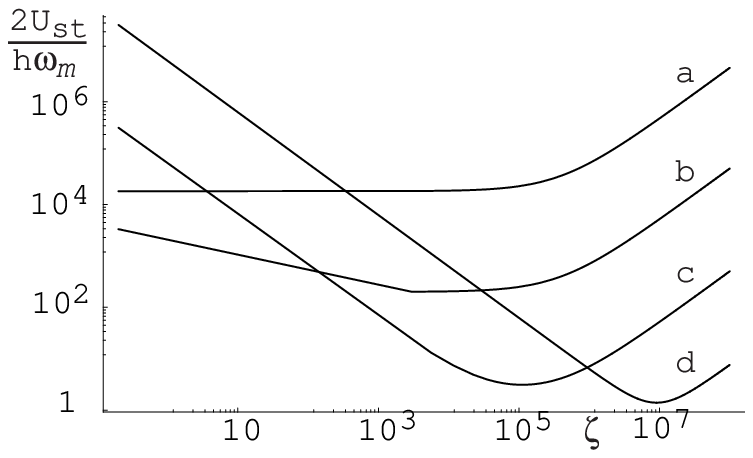}
\caption{
Rescaled steady-state energy 
$2U_{st}/\hbar \omega_{m}$ 
versus the rescaled input power $\zeta$, plotted for 
different values of $g_{1}$ (a: $g_{1}=10$, b: $g_{1}=10^{3}$,
c: $g_{1}=10^{5}$, d: $g_{1}=10^{7}$) at fixed ${\cal Q}=10^{7}$, 
and with $k_{B}T/\hbar \omega_{m}= 10^{5}$ and $\eta =0.8$.
The optimal input power $\zeta_{opt}$ correspondingly increases, and
for high gain values, ground state cooling can be 
achieved. }
\label{enesc-g}
\end{figure}

\begin{figure}[ht]
\includegraphics[width=3.5in]{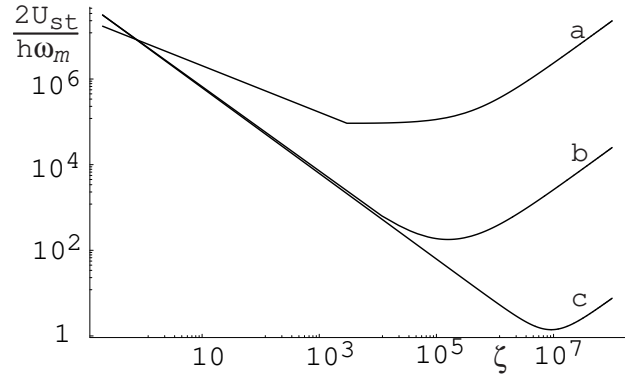}
\caption{
Rescaled steady-state energy $2U_{st}/\hbar \omega_{m}$ 
versus $\zeta$ for 
increasing values of the mechanical quality factor ${\cal Q}$ 
(a: ${\cal Q}=10^{3}$,
b: ${\cal Q}=10^{5}$, c: ${\cal Q}=10^{7}$) at fixed $g_{1}=10^{7}$,
and with $k_{B}T/\hbar \omega_{m}= 10^{5}$ and $\eta =0.8$. }
\label{enesc-q}
\end{figure}

The feedback scheme of Ref.~\cite{MVTPRL} 
has been introduced just for significantly cooling the 
cavity mirror. Let us therefore study the cooling capabilities
of this scheme and compare them with those of the cold damping scheme.
The stationary
oscillator energy $U_{st}$, neglecting the logarithmic 
correction of Eq.~(\ref{p2sc2})), can be written as
\begin{eqnarray}
&& U_{st} = \hbar \omega_m\left[\langle Q^{2}\rangle _{st}
+\langle P^{2}\rangle _{st}\right] = 
\frac{\hbar \omega_m}{8}\left[
\frac{g_{1}^{2}}{\eta \zeta }\frac{ 
\left(1+2{\cal Q}^{2}+g_{1}\right)}{\left(1
+g_{1}\right)\left({\cal Q}^{2}+g_{1}\right)} \right. \nonumber \\
&&+ \left.
\left(\zeta+ \frac{4k_B T}{\hbar \omega_{m}}\right)\frac{
\left(g_{1}^{2}+2{\cal Q}^{2}+g_{1}\right)}{\left(1
+g_{1}\right)\left({\cal Q}^{2}+g_{1}\right)}\right].
\label{ener}
\end{eqnarray} 
It is evident from Eq.~(\ref{ener}) that the effective temperature is
decreased only if both ${\cal Q}$ and $g_1$ are very large. At the same
time, the additional terms due to the feedback-induced noise and the 
back-action noise have to remain bounded for ${\cal Q} \to \infty$ and 
$g_1 \to \infty$, and this can be obtained by minimizing $U_{st}$ with
respect to $\zeta$ keeping ${\cal Q}$ and $g_1$ fixed. It is possible to
check that these additional terms are bounded only for very large ${\cal Q}$,
that is, if ${\cal Q}/g_1 \to \infty$ and in this case the minimizing
rescaled input power is $\zeta_{opt} \simeq g_1/\sqrt{\eta}$. Under these 
conditions, the steady state oscillator energy becomes
\begin{equation}
U_{st} \simeq \frac{\hbar \omega_m}{2}\left[\frac{1}{\sqrt{\eta}}+
\frac{2k_B T}{\hbar \omega_m}\frac{1}{g_1}\right]\;,
\label{enersimpl}
\end{equation}
showing that, in the ideal limit 
$\eta =1$, $g_1 \to \infty$, $\zeta \sim g_1 \to \infty$, ${\cal Q}/g_1 \to
\infty$, also momentum feedback is able to reach the
quantum limit $U_{st} =\hbar \omega_m /2$, i.e., it is able to cool
the mirror down to its quantum ground state. The behavior of the 
steady-state energy is shown in Figs.~\ref{enesc-g} and \ref{enesc-q}, 
where $U_{st}$ (in zero-point energy units $\hbar \omega_{m}/2$)
is plotted as a function of the rescaled input power $\zeta$. 
In Fig.~\ref{enesc-g}, $2U_{st}/\hbar \omega_{m}$ is plotted for 
increasing values of $g_{1}$ (a: $g_{1}=10$, b: $g_{1}=10^{3}$,
c: $g_{1}=10^{5}$, d: $g_{1}=10^{7}$) at fixed ${\cal Q}=10^{7}$, 
and with $k_{B}T/\hbar \omega_{m}= 10^{5}$ and $\eta =0.8$.
The figure shows the corresponding increase of the optimal input power 
minimizing the energy,
and that for high gain values, ground state cooling can be 
essentially achieved, even with a nonunit detection efficiency. 
In Fig.~\ref{enesc-q}, $2U_{st}/\hbar \omega_{m}$ 
is instead plotted for 
increasing values of the mechanical quality factor ${\cal Q}$ 
(a: ${\cal Q}=10^{3}$,
b: ${\cal Q}=10^{5}$, c: ${\cal Q}=10^{7}$) at fixed $g_{1}=10^{7}$.
The figure clearly shows the importance of ${\cal Q}$ in momentum
feedback and that ground state 
cooling is achieved only when ${\cal Q}$ is sufficiently large.
In the limit ${\cal Q}/g_1 \to \infty$, the steady state in the 
presence of momentum feedback
becomes very similar to that of cold damping, as it can be
checked by comparing the expressions for the steady state variances
in both cases. Under this limit, the two schemes have 
comparable cooling capabilities, even though it is evident that
cold damping is more suitable to reach ground state cooling just
because the requirement of a very good mechanical quality factor 
is not needed.
The possibility to reach ground
state cooling of a macroscopic mirror using the feedback scheme of 
Ref.~\cite{MVTPRL} was first pointed out, using
an approximate treatment, in \cite{RON}, where the need of a
very large mechanical quality factor is underlined. Here we confirm this
result using an exact QLE approach. 

The ultimate quantum limit of ground state cooling is achieved in 
both schemes only if {\em both} the input power and the feedback gain 
go to infinity. If instead the input power is kept fixed, the 
effective temperature does not monotonically decrease for increasing 
feedback gain, but, as it can be easily seen from Eqs.~(\ref{ener2})
and (\ref{ener}), there is an optimal feedback gain, giving a minimum steady 
state energy, generally much greater than the quantum ground state 
energy. The existence of an optimal feedback gain at fixed input 
power is a consequence of the feedback-induced noise term originating 
from the quantum input noise of the radiation. In a classical treatment 
neglecting all quantum radiation noises, one would have instead erroneously
concluded that the oscillator energy can be made arbitrarily small, by 
increasing the feedback gain, and independently of the radiation input 
power. This is another example of the importance of including the radiation 
quantum noises, showing again that a full quantum treatment is necessary
to get an exhaustive description of the system dynamics \cite{LETTER}.

The experimental achievement of ground state cooling via feedback is 
prohibitive with present day technology. For example, the experiments 
of Refs.~\cite{HEIPRL,PINARD,ATTO} have used feedback gains up to 
$g_{2}=40$ and an input power corresponding to $\zeta \simeq 1$, and 
it is certainly difficult to realize in practice the limit of very 
large gains and input powers. This is not 
surprising, since this would imply the preparation of a mechanical
macroscopic degree of freedom in its quantum ground state, which 
is remarkable.

\section{Nonclassical effects with momentum feedback}

The explicit dependence upon the mechanical quality factor
${\cal Q}$ of the steady state variances makes momentum feedback
less suitable than cold damping for ground state cooling. 
However, the presence of
an additional, in principle tunable, parameter, makes the
steady state in the presence of the
feedback scheme of Ref.~\cite{MVTPRL} richer, and capable of showing
interesting and unexpected nonclassical effects at the macroscopic level.

For example, a peculiar aspect of momentum feedback 
is its capability of 
inducing steady-state correlations between the position and the 
momentum of the mirror, i.e., the fact that $\langle QP+PQ 
\rangle_{st} \neq 0$. This correlation can be evaluated solving the QLEs
Eqs.~(\ref{QEQ1})-(\ref{QEQ2}), and one gets \cite{PRALONG}
\begin{equation}
\frac{\langle QP+PQ\rangle _{st}}{2} =
\left(\frac{\zeta}{8}+
\frac{k_BT}{2\hbar\omega_{m}}
\right)
\frac{g_{1}{\cal Q}}{(1+g_{1})({\cal Q}^{2}+g_{1})}
-\frac{g_{1}^{2}}{8 \eta \zeta}
\frac{{\cal Q}}{(1+g_{1})
({\cal Q}^{2}+g_{1})}\,.
\label{QPSS1} 
\end{equation}
Due to the linearization of the problem 
(see Eqs.~(\ref{QLE2L1})-(\ref{QLE2L3})), 
the steady state in the presence of momentum
feedback is a Gaussian state, 
which however is never exactly a thermal state, 
because it is always $\langle Q^{2}\rangle _{st} \neq 
\langle P^{2}\rangle _{st}$ and $\langle QP+PQ\rangle _{st} \neq 0$.
Its phase space contours are therefore ellipses, rotated by
an angle 
$$\phi = \frac{1}{2}\arctan\left[\frac{\langle QP+PQ\rangle _{st}}{\left(
\langle Q^{2}\rangle _{st} - \langle P^{2}\rangle _{st}\right)}\right]$$
with respect to the $Q$ axis. The steady state becomes approximately
a thermal state only in the limit of very large ${\cal Q}$ (and ${\cal Q}^2
\gg g_1$), as it can be seen from Eqs.~(\ref{q2sc2}), (\ref{p2sc2}) and
(\ref{QPSS1}). This thermal state approaches the quantum ground state of
the oscillating mirror when also the feedback gain and the input power
become very large.

\begin{figure}[ht]
\includegraphics[width=3.5in]{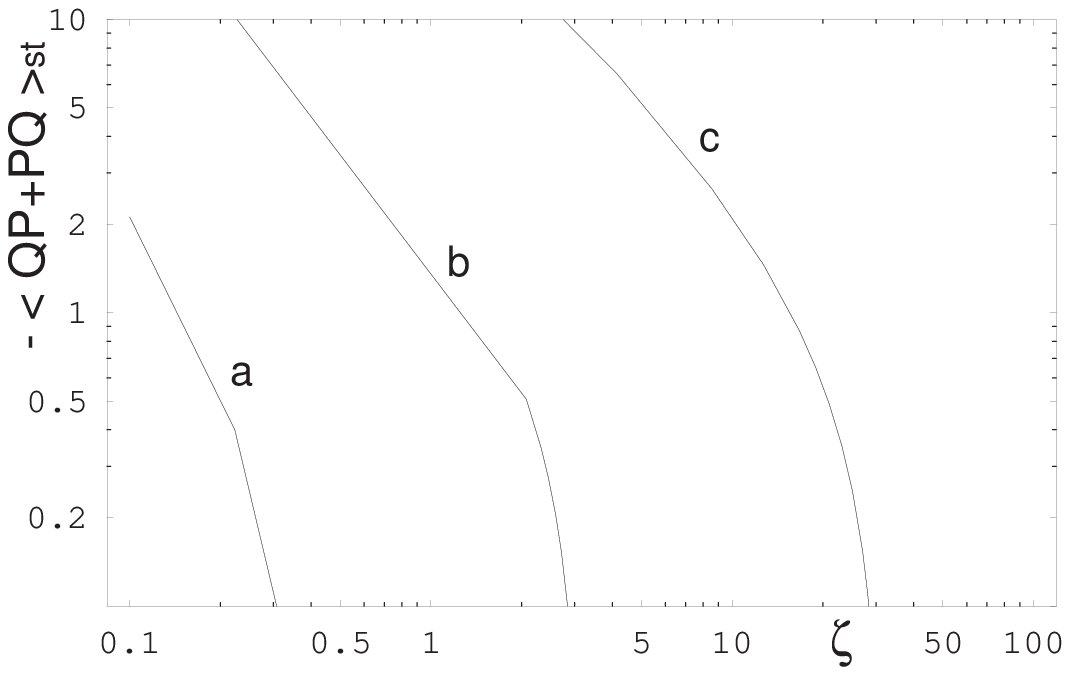}
\caption{ 
Steady state position-momentum correlation 
$-\langle QP+PQ \rangle _{st}$
versus $\zeta$ for 
three values of the feedback gain, $g_1=10^{5}$ (a),
$g_1=10^{6}$ (b) and $g_1=10^{7}$ (c). The other parameters are: 
${\cal Q}=10^{4}$, $k_{B}T/\hbar \omega_{m}=10^{5}$ and $\eta =0.8$.
}
\label{fig4}
\end{figure}

A first example of nonclassical behavior is that
this Gaussian steady state can become a
{\em contractive state}, which has been shown to be able
to break the standard quantum limit in \cite{YUEN}, when 
$\langle QP+PQ\rangle _{st}$ becomes negative, and this can be 
achieved at sufficiently large feedback gain, that is, when
$g_{1}>\eta \zeta \left(\zeta+ 4k_BT/\hbar\omega_{m}\right)$
(see Eq.~(\ref{QPSS1})). This is shown in Fig.~\ref{fig4}.

\begin{figure}[ht]
\includegraphics[width=3.5in]{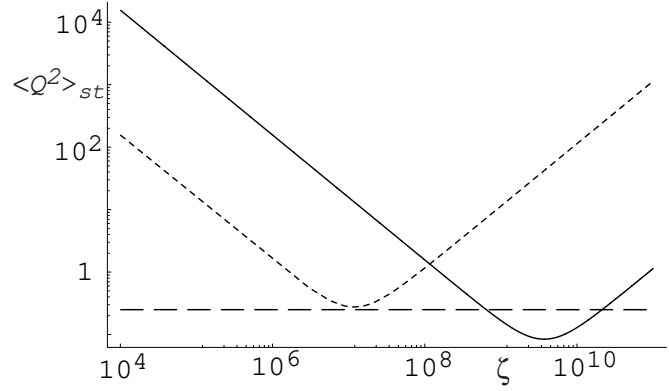}
\caption{ 
Steady state position variance $\langle Q^{2}\rangle _{st}$
versus $\zeta$ for 
two values of the feedback gain, $g_1=10^{7}$ (dotted line),
and $g_1=10^{9}$ (full line). The dashed
line denotes the standard quantum limit 
$\langle Q^{2}\rangle _{st}= 1/4$, while the other parameters are: 
${\cal Q}=10^{4}$, $k_{B}T/\hbar \omega_{m}=10^{5}$ and $\eta =0.8$. 
}
\label{squee}
\end{figure}

A second interesting example is that it is possible to achieve
steady state position squeezing, that is, to beat the standard quantum
limit $\langle Q^{2}\rangle _{st} <1/4$. The strategy is similar to that
followed for cooling. First of all one has to minimize 
$\langle Q^{2}\rangle _{st}$ with respect to the input power $\zeta$
at fixed $g_1$ and ${\cal Q}$, obtaining
\begin{equation}
\langle Q^{2}\rangle _{st}^{min}=\frac{g_1 {\cal Q}\sqrt{1+{\cal Q}^2+g_1}
}{4\sqrt{\eta}(1+g_1)({\cal Q}^2+g_1)}+\frac{k_B T}{2\hbar \omega_m}
\frac{{\cal Q}^2}{(1+g_1)({\cal Q}^2+g_1)}.
\label{q2min}
\end{equation}
This quantity can become arbitrarily small in the limit of very large
feedback gain, and provided that $g_1 \gg {\cal Q}^2$. That is, differently
from cooling, position squeezing is achieved in the limit $g_1 \to \infty$
(implying $\zeta \to \infty$), and there is no condition on 
the mechanical quality
factor. Under this limiting conditions, $\langle Q^{2}\rangle _{st}$ goes to
zero as $g_1^{-1/2}$, and, at the same time, $\langle P^{2}\rangle _{st}$
diverges as $g_1^{3/2}$, so that, in this limit, the steady state 
in the presence of momentum feedback approaches the position eigenstate
with $Q=0$, that is, the mirror tends to be perfectly localized at its
equilibrium position. The possibility to beat the standard quantum limit
for the position uncertainty is shown in Fig.~\ref{squee}, where
$\langle Q^{2}\rangle _{st}$ is plotted versus $\zeta$ for two different
values of the feedback gain, $g_1=10^7$ (dotted line), and $g_1=10^9$ (full 
line), with ${\cal Q}=10^4$, $k_B T/\hbar \omega_m= 10^5$, and 
$\eta =0.8$. For the higher value of the feedback gain, the standard quantum
limit $\langle Q^{2}\rangle _{st}= 1/4$ (dashed line)
is beaten in a range of values of the input power $\zeta$.

\begin{figure}[ht]
\includegraphics[width=3.5in]{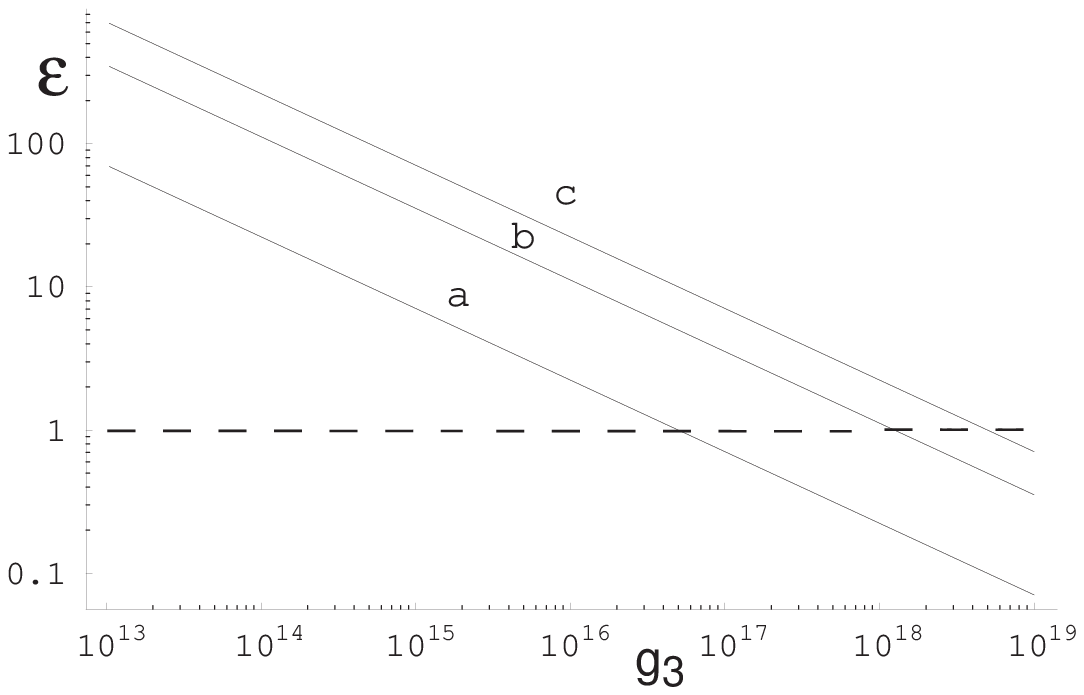}
\caption{
The marker of entanglement ${\cal E}$ is plotted 
versus the feedback gain $g_{1}$ for three values of 
the mechanical quality factor ${\cal Q}=10^{3}$ (a),
${\cal Q}=3\times 10^{3}$ (b) and ${\cal Q}=10^{4}$ (c).
The other parameters are: 
$k_{B}T/\hbar \omega_{m}=10^{5}$ and $\eta =0.8$.
}
\label{fig6}
\end{figure}

This capability to achieve quantum squeezing can be exploited to
get a third interesting and somewhat unexpected nonclassical effect,
i.e., the possibility to have an {\em entangled} stationary state
of the two movable mirrors in the ring cavity scheme of Fig.~2.
This entangled state corresponds to a squeezed state of the relative
motion of the two mirrors, in which their positions
are strongly correlated.
Such a squeezing could be obtained by applying the feedback scheme of 
Ref.~\cite{MVTPRL} to the relative motion of the mirror, which
amounts to apply a continuous position shift
to both mirror (see Fig.~2), in such a way that each mirror is shifted 
exactly opposite to the other. This means having the following additional
term in the QLE for a generic operator ${\cal O}(t)$ 
of the two-mirror system, given by
\begin{equation}\label{DOSTRA2}
\dot{{\cal O}}_{fb}(t)=
i\frac{\sqrt{\gamma_{c}}}{\eta}
\int_{0}^{t} dt'G_{mf}^{-}(t')Y_{out}(t-t')
\left[g_{mf}^{-}P_{-}(t),{\cal O}(t)\right]\,,
\end{equation}
where $G_{mf}^{-}(t)$ and $g_{mf}^{-}$ are the corresponding 
feedback transfer function and gain factor, respectively.  
Assuming again a zero-delay time limit $G_{mf}^{-}(t)=\delta(t)$,
the treatment of the preceding Section for a single movable mirror
can be easily applied also to the case of this feedback loop
acting simultaneously on the two mirrors. In fact,
redefining the corresponding rescaled,
dimensionless, input 
power of the driving laser 
\begin{equation} 
\tilde{\zeta} =  \frac{32 \tilde{G}^{2}\tilde{\beta} ^{2}}{\gamma_m\gamma_{c}}=
\frac{128\tilde{G}^{2}}{\hbar \omega_{0}\gamma_{m}\gamma_{c}^{2}}\wp ,
\label{zeta2}
\end{equation}  
and the corresponding rescaled feedback gain
$g_3 = -4\sqrt{2}\tilde{G}\tilde{\beta} g_{mf}^-/\gamma_{m}$, 
one gets an expression for the relative distance variance
$\langle Q_{-}^{2}\rangle _{st}$ analogous to that of Eq.~(\ref{q2sc2}),
that is,
\begin{equation}
    \langle Q_{-}^{2}\rangle _{st} = 
    \frac{g_{3}^{2}}{8 \eta \tilde\zeta} 
    \frac{1+{\cal Q}^{2}+g_{3}}{\left(1
    +g_{3}\right)\left({\cal Q}^{2}+g_{3}\right)}+
    \left[\frac{\tilde\zeta}{8}+
    \frac{k_B T}{2\hbar \omega_{m}}\right]
    \frac{{\cal Q}^{2}}{\left(1+g_{3}\right)
    \left({\cal Q}^{2}+g_{3}\right)},
    \label{Q-2st}
\end{equation}
Entanglement between the two mirrors in the steady state could be
demonstrated using one of the sufficient critera for entanglement
recently appeared in the literature 
\cite{TAN,DUAN,SIMON,PRL02,REID,NOI}.
The entanglement criterion easiest to satisfy is the so-called
product criterion \cite{TAN,PRL02,REID,NOI}, which, using the
oscillator operators defined above, can be written as
\begin{equation}\label{ent}
    {\cal E}=16\; \langle Q_{-}^{2} \rangle_{st}\;
    \langle P_{+}^{2} \rangle_{st} < 1\,,
\end{equation}
where we have defined the marker of entanglement ${\cal E}$. This means
that the two movable cavity mirrors become entangled when the above product
of variances of the relative motion and of the center of mass of the 
system of two mirrors goes below a certain limit. The total momentum
is not affected either by the radiation field, or by 
feedback (see Eq.~(\ref{DOSTRA2})), and therefore
it is simply given by the Brownian motion contribution 
$\langle P_{+}^{2} \rangle_{st}=\langle P_{+}^{2}\rangle _{BM}$, which is
\begin{equation}
    \langle P_{+}^{2}\rangle _{BM} = \frac{k_B T}{2\hbar \omega_{m}}
  + \frac{\gamma_{m}}{\pi \omega_{m}}\ln\left(\frac{\hbar \varpi}{2 \pi k 
    T}\right). \label{P+2st}
\end{equation}
In order to achieve the best conditions for entanglement
one follow the same strategy adopted to get position squeezing in the 
Fabry-Perot cavity case, because one has essentially to squeeze the
relative distance. One has 
first to minimize Eq.~(\ref{Q-2st}) with respect to the input power 
$\tilde\zeta$
at fixed $g_3$ and ${\cal Q}$, obtaining the optimal input power
\begin{equation}\label{ztopt}
    \tilde\zeta_{opt}=\frac{g_{3}}{\cal Q}\sqrt{
    \frac{1+{\cal Q}^{2}+g_{3}}{\eta}}\,,
\end{equation}
giving exactly expression (\ref{q2min}) for 
$\langle Q_{-}^{2}\rangle _{st}^{min}$, with $g_1 \to g_3$. 
This quantity can become arbitrarily small in the limit of very large
feedback gain (implying a very large input power, see Eq.~(\ref{ztopt}))
and fixed ${\cal Q}$. Since 
$\langle P_{+}^{2} \rangle_{st}$ does not depend upon $g_3$, 
this implies that the marker of entanglement ${\cal E}$ 
can be reduced below $1$. This means the possibility to entangle the 
two mirrors by only means of the feedback action.
This is shown in Fig.~\ref{fig6}. Notice that in such a case an 
extremely high feedback gain is needed, showing the difficulty in
obtaining such effect.
This is in apparent contrast with what has been obtained in 
Ref.~\cite{PRL02} showing that entanglement could be achieved
without feedback, and in a large temperature range,
in the frequency domain, within a narrow bandwidth around the mechanical 
resonance.
The key point is that in the present model, the feedback-induced 
entanglement is achieved for the full stationary state,
integrated over all frequencies, and is not
limited to a small bandwidth in the frequency domain.
This is a much stronger effect, and it is not surprising that it 
is much more difficult to achieve.

\section{Conclusions}

We have studied how quantum feedback schemes can be used in 
optomechanical system to achieve cooling of vibrational degrees of 
freedom of a mirror.
We have analysed and compared the momentum feedback scheme introduced in
Ref.~\cite{MVTPRL}, and the cold damping scheme of 
\cite{HEIPRL,PINARD,ATTO,COURTY}. 
The main effect of feedback is the increase of mechanical damping, 
accompanied by the introduction of a controllable, measurement-induced, 
noise. The increase of damping means reduction of the susceptibility 
at resonance, and the consequent suppression of the 
resonance peak in the noise spectrum. 
We have then shown the possibility 
to achieve the ultimate quantum limit of ground state cooling
with both feedback schemes.
Cold damping is more suitable for cooling than momentum feedback
because its capabilities are not influenced by the mechanical quality
factor of the oscillators ${\cal Q}$. 
Instead momentum feedback requires a very large 
${\cal Q}$, and in this limit it tends to coincide with cold damping.
However, for fixed ${\cal Q}$ and for very large feedback gain 
(and input powers), the steady state in the presence of momentum feedback
shows unexpected nonclassical features.
In fact, it can become a position squeezed, or contractive state.
Moreover, applying momentum feedback to the relative motion of two
movable mirrors of a ring cavity, one can even get an entangled stationary
state of the two mirrors, again in the limit of very large feedback
gains and input powers.

Both ground state cooling and nonclassical effects are achieved in a
parameter region which is extremely difficult to achieve with present
day technology. This is expected, because it would imply the demonstration
of genuine quantum effects for a macroscopic mechanical degree of freedom.

\newpage

\newpage

\newpage

\end{document}